\documentclass{article}
\usepackage{epsfig,amssymb}
\newcommand{\be}{\begin{equation}}
\newcommand{\ee}{\end{equation}}
\newcommand{\ba}{\begin{array}{c}}
\newcommand{\ea}{\end{array}}
\newcommand{\bea}{\begin{eqnarray}}
\newcommand{\eea}{\end{eqnarray}}

\begin{document}

\date{\today}
\title{ Extracting Classical Correlations from a Bipartite Quantum System}
\author{ S. Hamieh, J. Qi, D. Siminovitch, and M. K. Ali }
\maketitle
{\centerline{ Department of Physics, University of Lethbridge,
}}\centerline{4401 University Drive, Lethbridge, AB, T1K 3M4, Canada}

\begin{abstract}
In this paper we discuss the problem of splitting the  total 
correlations for a bipartite quantum state described by
the Von Neumann mutual information into classical and
quantum parts. We propose a measure of the classical
correlations as the difference between the Von
Neumann mutual
information and the relative entropy of entanglement. We
compare this measure with
different measures proposed in the literature.

\end{abstract}
\vspace{1cm}
\newpage

In quantum information theory any bipartite quantum  state
can be used as a quantum channel to send quantum 
information from sender to receiver \cite{Benn93}. If 
the  state is maximally entangled then the fidelity
of  the channel is one, however if the state is not maximally
entangled and/or mixed, the degree of success is less than one.
 From this point of view it 
is interesting to  quantify the quantum and the 
classical correlations in a quantum state. Different
measures of the quantum correlations
 have been proposed in the past few years.
Bennett et al. \cite{Benn96} proposed entanglement of formation and 
distillable entanglement, where the former is the 
number of the maximally entangled states that is necessary to  create
the state by local operations and classical communication
(LOCC), and the latter is the number of maximally entangled
states that can be distilled from the state by LOCC.
Vidal and Werner \cite{Vida01} proved that the negativity
defined as two times the absolute values of the negative
eigenvalue of the partial transpose of a state is an 
entanglement monotone and therefore is a good entanglement measure.
In this paper we will use the relative entropy of entanglement \cite{Vedr97}
as a  measure of entanglement. All these 
entanglement measures 
are proved to be consistent with each other, however
for the  known measures of classical correlations this is
not proved to be true \cite{Vedr01}. 

The first measure of  the classical correlations of a bipartite
 state $\rho_{AB}$ was proposed in
 \cite{Vedr97} as a distance between the nearest separable
state $\rho^*$ of $\rho_{AB}$ and $\rho^{*}_{A}\otimes\rho^{*}_{B}$
that is 
\be C_1(\rho_{AB})=S(\rho^{*}||\rho^{*}_{A}\otimes\rho^{*}_{B})\,,\ee
 where  $\rho_A^*={\rm Tr}_{B}(\rho^*)$ and
  $\rho_B^*={\rm Tr}_{A}(\rho^*)$  are the  reduced density operators 
of $\rho^*$ and 
 $S(\rho^*||\rho^*_A\otimes\rho^*_B)= {\rm Tr}\{\rho^*[\log_2
(\rho^*)-\log_2(\rho^*_A\otimes\rho^*_B)]\}$ is the relative entropy  
of $\rho^*$ with respect to 
$\rho^*_A\otimes\rho^*_B$. Here the  relative entropy is
used as distance \footnote{ This distance  is not 
a  metric because it is not symmetric under interchange of the
states}. Note that this measure of the classical correlations,
 and similar  measures  
based on the distance between  $\rho^{*}$ and
 $\rho_{A}\otimes\rho_{B}$ such  as \cite{Vedr01}
 \be C_2(\rho_{AB})=S(\rho^{*}||\rho_{A}\otimes\rho_{B})\,,\ee
 are
equal to  the Von Neumann
mutual information 
\be I(\rho_{AB})=S(\rho_A)+S(\rho_B)-S(\rho_{AB})=
S(\rho_{AB}||\rho_A\otimes\rho_B)\,,\ee
for a  set of separable states.  Recently  
Vedral et al. \cite{Vedr01} proposed a measure of the classical correlations
based on the maximum information that could be  extracted
  on a
subsystem $B$ of $\rho_{AB}$  
  by making  a measurement on the other
subsystem $A$  
so
 \be \chi_{B}=\max_{A_iA_i^{\dag}} S(\rho_B)- \sum_i p_iS(\rho_B^i)\,,\ee

where 
$\rho_B^i={{\rm Tr}_A A_i\rho_{AB}A_i^{\dag}/ {\rm Tr}[{\rm Tr}_A
(A_i\rho_{AB}A_i^{\dag})]}$ and the maximum is taken over all
POVM measurements that $A$ can perform for his state such
$\sum_iA_iA_i^{\dag}=I$.
 This measure  is   not proved to 
be symmetric under interchange of 
the  subsystems $A$ and $B$ and even if it could
increase under LOCC that should be expected for  
classical correlations.  

In this paper we suggest a new measure of the classical
correlation that is symmetric under the interchange of $A$ and $B$
and we compare it with Vedral et al. measures \cite{Vedr97,Vedr01}
for the class of all Werner states.

We define the classical correlations as 
the difference between the total correlations
measured by the Von Neumann mutual information   
and the quantum correlations measured by the relative entropy of entanglement 
 \be \Psi(\rho_{AB})=S(\rho_{AB} || \rho_A\otimes\rho_B)-
\min_{\rho^*\in {\cal D}} S(\rho_{AB} || \rho^*) \,,\ee
where          
${\cal D}$ is the set of all separable states in the Hilbert space
on which $\rho_{AB}$ is defined.
$ \Psi(\rho_{AB})$ satisfies some properties that we  expect
for the classical correlations:
\begin{enumerate}

\item By definition $\Psi(\rho_{AB})$ is symmetric under interchange
of $A$ and $B$ as expected.
\item $\forall$ $\rho_{AB}$ we have
 $\Psi(\rho_{AB})\geq 0$ because 
the relative entropy of entanglement is the minimum over the set of
separable states including  $\rho_A\otimes\rho_B$.
\item  $\Psi(\rho_{AB})$ is invariant under local unitary transformations
due to the fact that the relative entropy and ${\cal D}$
 are invariant under
this transformation \cite{Vedr97}. 
\item $\Psi(\rho_{AB})=0$ iff $\rho_{AB}=\rho_{A}\otimes \rho_{B}$
(evident from the definition). This condition
 is a natural requirement
as in this case $A$ and $B$ are completely separable so
there are no expected classical correlations
between them.
\item  $\Psi(\rho_{AB})$ could increase under LOCC due to an increase in mutual
information, while the relative entropy of entanglement
cannot. This is an important requirement that enables  the enhancement
of classical correlations by classical communication.
\item For separable states,  $\Psi(\rho_{AB})$ cannot increase
by local operations only because in this case,  $\Psi(\rho_{AB})$ is just 
equal to the mutual information
\cite{Vedr97}. We do not prove  that  in general however it
is natural to conjecture it  because the decreasing part 
of the total correlations  should include the decreasing part of the
quantum correlations.
\end{enumerate}
This measure of the classical correlation is  superadditive 
in the sense that  $\Psi(\rho\otimes\rho)\geq 2\Psi(\rho)$ due
to the fact that the mutual information is additive, whereas 
the relative entropy of entanglement is subadditive.
We will now compare  $\Psi(\rho_{AB})$ with the measures proposed
by Vedral et al. \cite{Vedr97,Vedr01}.  

\begin{figure}[tb]
\centerline{\psfig{width=10.5cm,figure=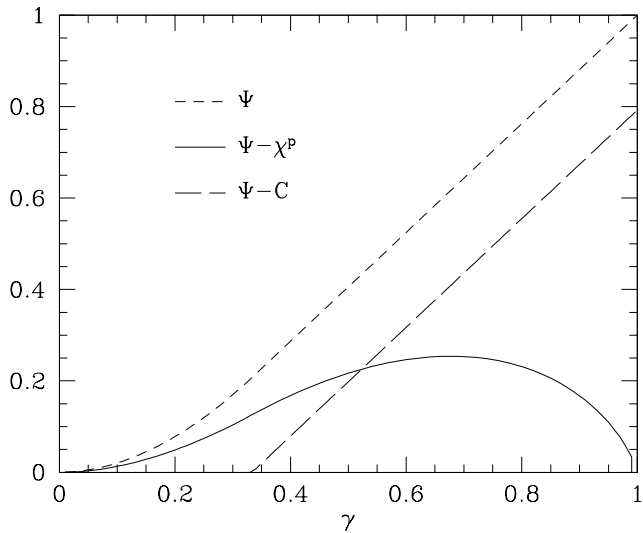}}
\vspace{-2.2cm}
\caption{ Classical correlations for the class of Werner states
measured by $\Psi(\rho_{AB})$ (dotted line), 
 $\Psi(\rho_{AB})-\chi_{B}^p$ (solid line)
and $\Psi(\rho_{AB})-C$ (dashed line).
\label{fig1} }
\end{figure} 

 For pure states,  all the measures of the classical
correlations  including our measure
are equal  to the Von Neumann entropy 
of the subsystems $A$ or $B$ 
$S(\rho_B)=S(\rho_A)$.

Note that, 
 according to our proposed
definition of the classical correlation
all the correlations contained in separable states are
classical. On the other hand, recently it has turned out that some of
them can be still of quantum nature 
see Ollivier and Zurek 
\cite{Zur01}). This however does {\it not} nullify our result.
 Since there is no hope for unique measure of the classical
correlations, some operational measures  take into account {\it
some aspects} of the correlations and others 
may take different aspects. In a similar
way, distillable entanglement  vanishes on some entangled states (ie.
bound entangled ones) but turns out to point out all the entanglement
that is directly useful in quantum information theory.

 To compare these measures for mixed states,  we
choose as an example the class of all 
Werner states in $2\otimes 2$  
\be \rho_{AB}={(1-\gamma)\over 4}I_4 +\gamma |\psi^+\rangle\langle\psi^+
|\,,\ee
which interpolates between the unpolarized state $I_4/4$ and the
Bell state
$|\psi^{+}\rangle={|00\rangle+|11\rangle \over \sqrt{2}}$
as $\gamma$ varies between 0 and 1.

In figure (\ref{fig1}) 
 we show the classical correlations measured by $\Psi(\rho_{AB})$.
To compare we plot also in this figure 
 $\Psi(\rho_{AB})-\chi_{B}^p$ (
the index $p$ introduced  to denote that the measurement is
  projective
\footnote{In this case, the measurement
 is not optimized for all von Neumann measurement
but calculated for some fixed one, namely the projective measurement.}) and
 $\Psi(\rho_{AB})-C$ (in this case
$C_1(\rho_{AB})=C_2(\rho_{AB})=C$). From 
figure (\ref{fig1}) we can see that $\Psi(\rho_{AB})-\chi_{B}^p$ is always
positive so it remains  an open question to determine
if there exists 
a POVM measurement which could make   $\Psi(\rho_{AB})-\chi_{B}=0$
or even negative. However, it is clear from  figure (\ref{fig1}) 
that  $\Psi(\rho_{AB})$ is different from $C_1(\rho_{AB})$
and $C_2(\rho_{AB})$ (this is a counterexample). We can also conclude from  
  figure (\ref{fig1}) that $\chi_B$ cannot be equal to 
$C_1(\rho_{AB})$
or $C_2(\rho_{AB})$ as $\chi_B>\chi_B^p>C$ for $\gamma>0.53$.

In conclusion, in  this paper we have investigated the problem of extracting
the classical correlations from bipartite quantum systems. We
have  discussed
different measures of the classical correlations and we have suggested
a measure based on the difference between the mutual
information and the relative entropy of entanglement.
We have compared different measures of the classical
correlations with our measure using the class of Werner states as a specific
example. Although we have proved from this
example that neither $\Psi(\rho_{AB})$ nor $\chi_B$ could be equal to
$C_1(\rho_{AB})$ or $C_2(\rho_{AB})$, it remains an open question whether
  $\Psi(\rho_{AB})$ is equal to  $\chi_B$ or not. 
Also 
remaining  as an open question is whether or not  
 $\Psi(\rho_{AB})$ 
is non increasing
 under local operations in the case
when the bipartite quantum state is non separable.
We hope to address some of these issues in the near future.

\end{document}